\documentclass[oldversion]{aa}

\usepackage{epsfig}
\usepackage[switch]{lineno}
\usepackage{graphicx}
\usepackage[varg]{txfonts}
\usepackage{natbib}
\renewcommand{\tabcolsep}{0.7mm}
\bibpunct{(}{)}{;}{a}{}{,} 

\begin{document}

\title{Evidence of heavy-element ashes in thermonuclear X-ray bursts
  with photospheric superexpansion}

\titlerunning{Superexpansion X-ray bursts}
\authorrunning{in 't Zand and Weinberg}

\author{J. J. M.~in~'t~Zand\inst{1} \& N. N. Weinberg\inst{2}}

\offprints{J.J.M. in 't Zand, email {\tt jeanz@sron.nl}}

\institute{     SRON Netherlands Institute for Space Research, Sorbonnelaan 2,
                3584 CA Utrecht, the Netherlands
           \and
                Astronomy Department and Theoretical Astrophysics Center,
                University of California, Berkeley, 601 Campbell Hall,
                Berkeley, CA 94720, U.S.A.
          }

\date{\em Accepted for publication on April 28, 2010}

\abstract{A small subset of thermonuclear X-ray bursts on neutron
  stars exhibit such a strong photospheric expansion that for a few
  seconds the photosphere is located at a radius $r_{\rm ph} \ga 10^3
  \textrm{ km}$.  Such `superexpansions' imply a large and rapid
  energy release, a feature characteristic of pure He burst
  models. Previous calculations have shown that during a pure He
  burst, the freshly synthesized heavy-element ashes of burning can be
  ejected in a strong radiative wind and produce significant spectral
  absorption features.  We search the burst data catalogs and
  literature and find 32 superexpansion bursts, 24 of which were
  detected with BeppoSAX and three with RXTE at high time
  resolution. We find that these bursts exhibit the following
  interesting features: (1) At least 31 are from (candidate)
  ultracompact X-ray binaries in which the neutron star accretes
  hydrogen-deficient fuel, suggesting that these bursts indeed ignite
  in a helium-rich layer.  (2) In two of the RXTE bursts we detect
  strong absorption edges during the expansion phase. The edge
  energies and depths are consistent with the H-like or He-like edge
  of iron-peak elements with abundances $\ga100$ times solar,
  suggesting that we are seeing the exposed ashes of nuclear burning.
  (3) The superexpansion phase is always followed by a moderate
  expansion phase during which $r_{\rm ph} \sim 30\textrm{ km}$ and
  the luminosity is near the Eddington limit. (4) The decay time of
  the bursts, $\tau_{\rm decay}$, ranges from short ($\approx
  10\textrm{ s}$) to intermediate ($\ga 10^3 \textrm{ s}$). However,
  despite the large range of $\tau_{\rm decay}$, the duration of the
  superexpansion is always a few seconds, independent of $\tau_{\rm
    decay}$. By contrast, the duration of the moderate expansion is
  always of order $\tau_{\rm decay}$. (5) The photospheric radii
  $r_{\rm ph}$ during the moderate expansion phase are much smaller
  than steady state wind models predict. We show that this may be
  further indication that the wind contains highly non-solar
  abundances of heavy elements.
  
}

\maketitle

\section{Introduction}
\label{intro}

X-ray bursts are thermonuclear shell flashes that result from the
unstable burning of fuel accreted on the surface of a neutron star
(NS) from a Roche-lobe filling companion star
(\citealt{gri76,woo76,mar77,lam78}; for reviews, see \citealt{lpt95,
  stroh06}). Approximately $20 \%$ of all bursts exhibit some degree
of photospheric radius expansion (PRE; see, e.g., \citealt{gal08}) due
to a flux exceeding the Eddington limit, resulting in a radiatively
driven expansion of the photosphere \citep{lew84, ebi83, taw84a,
  taw84b}. During the expansion, the emission area increases while the
bolometric luminosity remains nearly constant (near the Eddington
limit). As a result, the photosphere cools and the detected X-ray flux
decreases. Eventually the bolometric luminosity decreases, the
photosphere returns to the NS surface, and the X-ray flux rises again,
signifying the start of the main burst.

In a small fraction of bursts (we estimate no more than a few tenths
of a percent), the PRE is so extreme that for a few seconds the
detected X-ray flux decreases to less than 1\% of its peak value. Such
extreme PREs create the illusion of a precursor to the main burst, as
first reported by \cite{hof78}. Given that the peak temperature of a
burst is typically $2.5\textrm{ keV}$, close to the soft end of a
typical X-ray device's bandpass, such a dip implies that the effective
temperature decreases to less than $0.2 \textrm{ keV}$ (assuming pure
black body radiation at a constant bolometric flux and a hydrogen
column density of $10^{22}\textrm{ H-atoms cm}^{-2}$, typical for an
X-ray burst source). This corresponds to a radius expansion factor of
$100$ or more, implying that the photosphere moves out to radii
$r_{\rm ph} \ga 10^3 \textrm{ km}$. We call such an extreme PRE a {\em
  superexpansion} to distinguish it from the much more common {\em
  moderate} PRE with an expansion factor of a few.

In X-ray burst models, bursts that ignite in helium-rich layers have
high peak luminosities that often exceed the Eddington limit and drive
strong radiative winds \citep{han82, Paczynski:83, Fujimoto:87,
  woo04}.  Superexpansion bursts are the most luminous of all X-ray
bursts and may be the result of nearly pure helium ignition; as such
they serve as a useful test bed for helium burst models. In addition,
calculations show that heavy-element ashes of nuclear burning can be
ejected in the winds of PRE bursts and produce strong spectral
absorption features \citep{wei06a}. The brightness and powerful,
long-lasting winds of superexpansion bursts makes them ideal
candidates for detecting spectral evidence of heavy-element ash
ejection.

Helium bursts can be of either short or long duration depending on the
properties of the accretion \citep{Fujimoto:81, bil98}. If the NS
accretes a H/He mixture, the bursts tend to be short (seconds). If,
however, the system is an ultracompact X-ray binary (UCXB) and the NS
accretes hydrogen deficient material, the bursts tend to be much
longer (several tens of minutes, i.e., the `intermediate-duration
bursts'; \citealt{zan05a,cum06}). This is because in UCXBs, unlike in
mixed H/He accretors, there is no stable CNO burning to keep the
freshly accreted envelope hot. As a result, the helium fuel ignites
at a greater depth, where the thermal time is longer. As we will show,
the superexpansion bursts that we find are all associated with (candidate)
UCXBs.

The volume of data on superexpansion has increased substantially since
the last detailed publication on extreme PREs \citep{jvp90}. At that
time only 4 superexpansion bursts were known. It therefore makes sense
to revisit the subject. In \S~\ref{obs} we report on an additional 28
superexpansion bursts, most of them found in archival BeppoSAX and
RXTE data. In \S~\ref{sec:pcaspec} we examine the spectroscopic
features of the best-covered superexpansion bursts. We find evidence
for a variable absorption edge that may be due to exposed
heavy-element ashes of nuclear burning.  In
\S~\ref{sec:spectra_nonPCA} we describe the spectroscopic features of
the other superexpansion bursts. We find that in all 32 bursts the
superexpansion phase is followed by a moderate expansion phase. In
\S~\ref{sec:texpansion} we show that the duration of the two phases
are uncorrelated.  We place these results in the context of models of
helium-rich X-ray bursts and radiation-driven NS winds in
\S~\ref{sec:discussion} and conclude in \S~\ref{sec:conclusions}.

\section{Search for superexpansion bursts} 
\label{obs}

\begin{figure}[t]
\includegraphics[width=\columnwidth,angle=0]{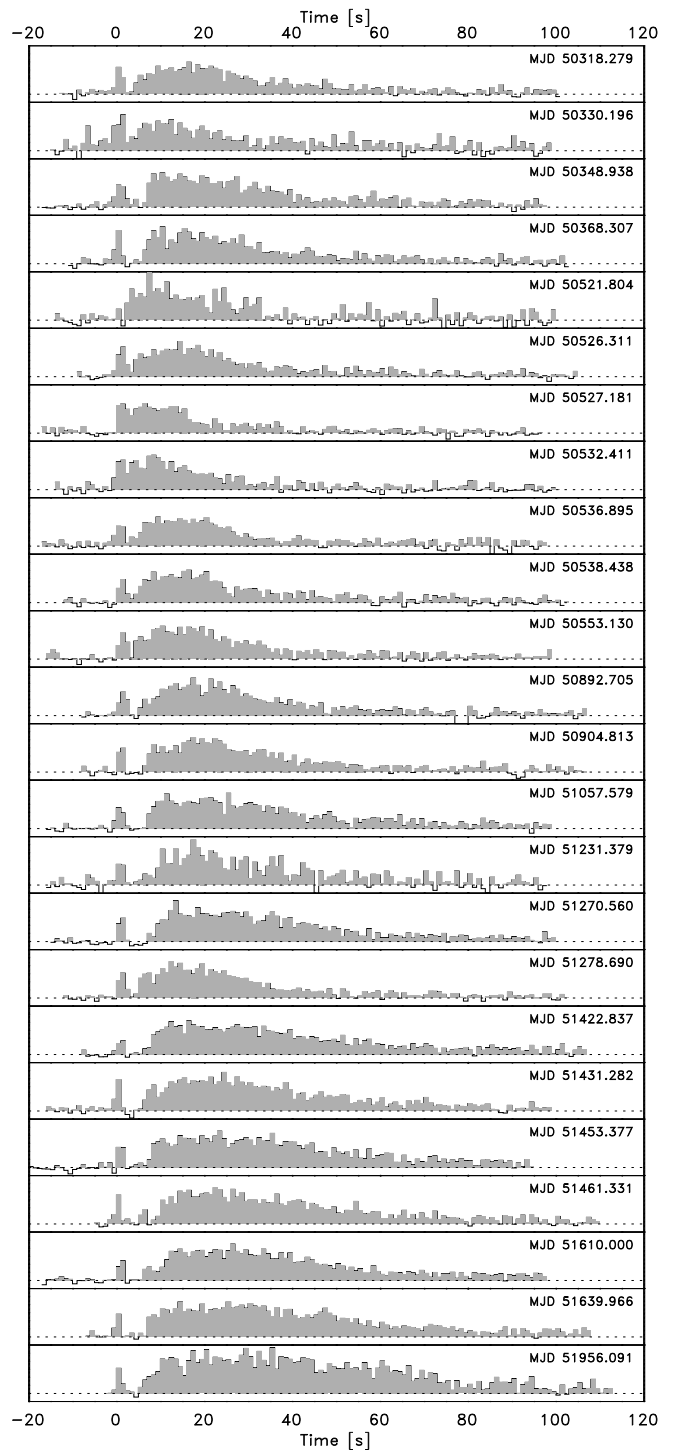}
\caption{Time profiles of the 2-30 keV photon flux of 24 X-ray bursts
  detected from 4U 1722-30 with BeppoSAX WFC (this includes 21 bursts
  with precursor, see Table~\ref{tab1}, and three without at MJD
  50521.804, 50527.181 and 50532.411). The photon flux is scaled
  between -1 and 4 phot~s$^{-1}$cm$^{-2}$. The horizontal dotted lines
  indicate a flux level of 0 phot~s$^{-1}$cm$^{-2}$. The MJD indicated
  in each panel refers to $t=0$~s. The profiles have been aligned to
  the start time of the precursor.
\label{figtz2wfc}}
\end{figure}

We searched for bursts with precursors in the literature
(\citealt{gal08,kuu03} and other relevant literature; see
Table~\ref{tab1}) and data of the Proportional Counter Array (PCA;
operational since January 1996) on RXTE up to and including 2008 and
in all data of the two Wide Field Cameras (WFCs; operational between
June 1996 and April 2002) on BeppoSAX. Both instrument packages
consist of proportional counters with similar efficiency curves and
spectral resolution (20\% full-width at half maximum), and bandpasses
that start at 2 keV. The PCA \citep{jah06} consists of 5 proportional
counter units (PCUs) that each have a propane-filled and a
xenon-filled gas layer. The xenon layer is the main detector and is
read out at high time and energy resolution between 2 and 60 keV. The
propane layer is used for anti-coincidence. It has only 0.125~s time
resolution and no energy resolution. However, since it has additional
sensitivity between 1.5 and 2 keV, it is useful for studying the soft
phases of X-ray bursts. The net collecting area of the xenon layers
combined is about 8000~cm$^2$, the field of view 2\degr\ (full-width
to zero response) and there is no angular resolution. The time
resolution can be set to 1~$\mu s$ at best. The two WFCs \citep{jag97}
each had a net collecting area of 140 cm$^2$, a field of view of
40\degr, an angular resolution of 5\arcmin\ in a bandpass of 2 to 28
keV. The time resolution was 0.5 ms. The observation program of the
PCA is in large part dedicated to bursting NSs, yielding large
exposure times \citep[of order megaseconds;][]{gal08} despite the
small field of view. Thanks to the wide field of view of the WFCs,
long exposure times were acquired on the majority of bursting NSs,
resulting in an unprecedented ability to detect rare and unexpected
events such as superexpansion bursts.

\begin{table}[ht]
\caption{Measurements of 32 bursts with superexpansion.\label{tab1}}
\begin{tabular}{l|c|c|c|c|r}
\hline\hline
   & \multicolumn{4}{|c|}{Duration (s)$^\diamond$} &  \\
              \cline{2-5}  
Instr./MJD             &           & Super     & Moderate  &  &Ref.$^\ddag$ \\
             & Precursor & expansion & expansion & \multicolumn{1}{c|}{$\tau_{\rm decay}^\P$} & \\
             & phase     & phase $t_{\rm se}$   & phase $t_{\rm me}$    &  & \\
\hline
\multicolumn{6}{c}{\em 4U 0614+09} \\
\hline
RXTE/51944.903    & 0.05 &  1.2 &  100  &   40(2) &  1 \\
\hline
\multicolumn{6}{c}{\em A 1246-588} \\
\hline
WFC/50286.290     &  3.0 &  1.5 &  54   &  38(5)  &  2 \\
WFC/51539.874     &  1.5 &  6.0 &  25   &  19(1)  &  2 \\
\hline
\multicolumn{6}{c}{\em 4U 1708-23 (probably)} \\
\hline
SAS-C/43181.834   &  4.2 &  6.0 &  304  & 300(50) &3,4 \\
\hline
\multicolumn{6}{c}{\em XB 1715-321} \\
\hline
SAS-C/42957.620   &  2.4 & 1.5  &  36   &   30$^\Delta$   &  3 \\
Hakucho/45170.231 &  3   & 4.0  &  105  &   85(5) &  5 \\
\hline
\multicolumn{6}{c}{\em 4U 1722-30} \\
\hline
WFC/50318.279     &  4.0 &  2.0 &   16  &  18(3)  &  6 \\
WFC/50330.196     &  2.0 &  5.0 &   20  &  16(6)  &  6 \\
WFC/50348.938     &  3.0 &  3.0 &   15  &  16(5)  &  6 \\
WFC/50368.307     &  2.0 & 5.0  &   14  &  19(3)  &  6 \\
WFC/50526.311     &  1.0 & 4.0  &   15  &  21(6)  &  6 \\
WFC/50536.895     &  2.5 & 3.5  &   11  &   9(3)  &  6 \\
WFC/50538.439     &  3.0 & 1.0  &   15  &  19(6)  &  6 \\
WFC/50553.130     &  2.5 & 1.5  &   17  &  12(2)  &  6 \\
WFC/50892.706     &  3.0 & 2.5  &   18  &  15(2)  &  6 \\
WFC/50904.813     &  2.5 & 4.0  &   21  &  19(3)  &  6 \\
WFC/51057.579     &  1.5 & 3.0  &   22  &  19(4)  &  6 \\
WFC/51231.379     &  2.0 & 5.0  &   17  &  26(11) &  6 \\
WFC/51270.560     &  2.0 & 5.5  &   25  &  14(2)  &  6 \\
WFC/51278.690     &  2.5 & 2.5  &   15  &  13(1)  &  6 \\
WFC/51422.838     &  2.0 & 6.5  &   22  &  28(5)  &  6 \\
WFC/51431.282     &  4.0 & 5.0  &   27  &  26(5)  &  6 \\
WFC/51453.377     &  1.5 & 5.5  &   20  &  24(6)  &  6 \\
WFC/51461.331     &  4.0 & 3.5  &   19  &  23(4)  &  6 \\
WFC/51610.000     &  3.0 & 3.5  &   20  &  17(3)  &  6 \\
WFC/51639.966     &  1.5 & 5.0  &   32  &  23(5)  &  6 \\
WFC/51956.091     &  2.0 & 3.5  &   41  &  29(10) &  6 \\
RXTE/50395.292    &  3.6 & 1.6  &   23  &  30.2(0.1) &  7 \\
\hline
\multicolumn{6}{c}{\em SLX 1735-269} \\
\hline
I'GRAL/52897.733  &  2.0 & 7.0  &   482 & 600(100)&  8 \\
\hline
\multicolumn{6}{c}{\em 4U 1820-30} \\
\hline
RXTE/51430.074$^\times$    & 15.0 & 2.3  &   1400& 2500    & 9 \\
\hline
\multicolumn{6}{c}{\em M15 X-2} \\
\hline
Ginga/47454.730   &  1.5 &  5.5 &   88  &   60    & 10 \\
WFC/51871.593     &  1.5 & 7.5  &   169 &155(11)  &    \\
\hline\hline
\end{tabular}

\noindent 
$^\diamond$See Fig. \ref{figdiagram} for a plot of $t_{\rm se}$ and
$t_{\rm me}$ vs. $\tau_{\rm decay}$.  $^\P$Numbers in parentheses
represent 1$\sigma$ uncertainties; $^\ddag$ 1 - \cite{kuu09}, 2 -
\cite{zan08}, 3 - \cite{hof78}, 4 - \cite{lew84}, 5 - \cite{taw84b}, 6
- \cite{kuu03}, 7 - \cite{mol00}, 8 - \cite{mol05}, 9 -
\cite{stroh02a}, 10 - \cite{jvp90}; $^\Delta$This number is rather
uncertain due to incomplete coverage of the burst; $^\times$superburst
\end{table}

Our search through $\approx 3400$ bursts and the literature turned up
32 superexpansion bursts from 8 different systems, see
Table~\ref{tab1}.  \emph{All the bursts that have been unambiguously
  identified with an object (7 out of 8 systems) are from (candidate)
  UCXBs} as listed by \cite{zan07}. The UCXB candidacy of the only
transient, XB 1715-321, has been put forward in \cite{jon07}. The one
burst not unambiguously identified with an object is from 4U 1708-23.
Of the 32 bursts listed, 22 are from 4U 1722-30 (in the globular
cluster Terzan 2), 21 of which were detected with BeppoSAX-WFC
\citep[see Fig.~\ref{figtz2wfc} and][]{kuu03}. Almost all bursts from
4U 1722-30 have a precursor; for a burster with precursors, it has
relatively frequent bursts (once every few days).  The superexpansion
burst from 4U 1820-30 is a superburst \citep[][hereafter
  SB02]{stroh02a}.

\section{Spectral analysis of the 3 PCA bursts}
\label{sec:pcaspec}

\begin{figure*}[t]
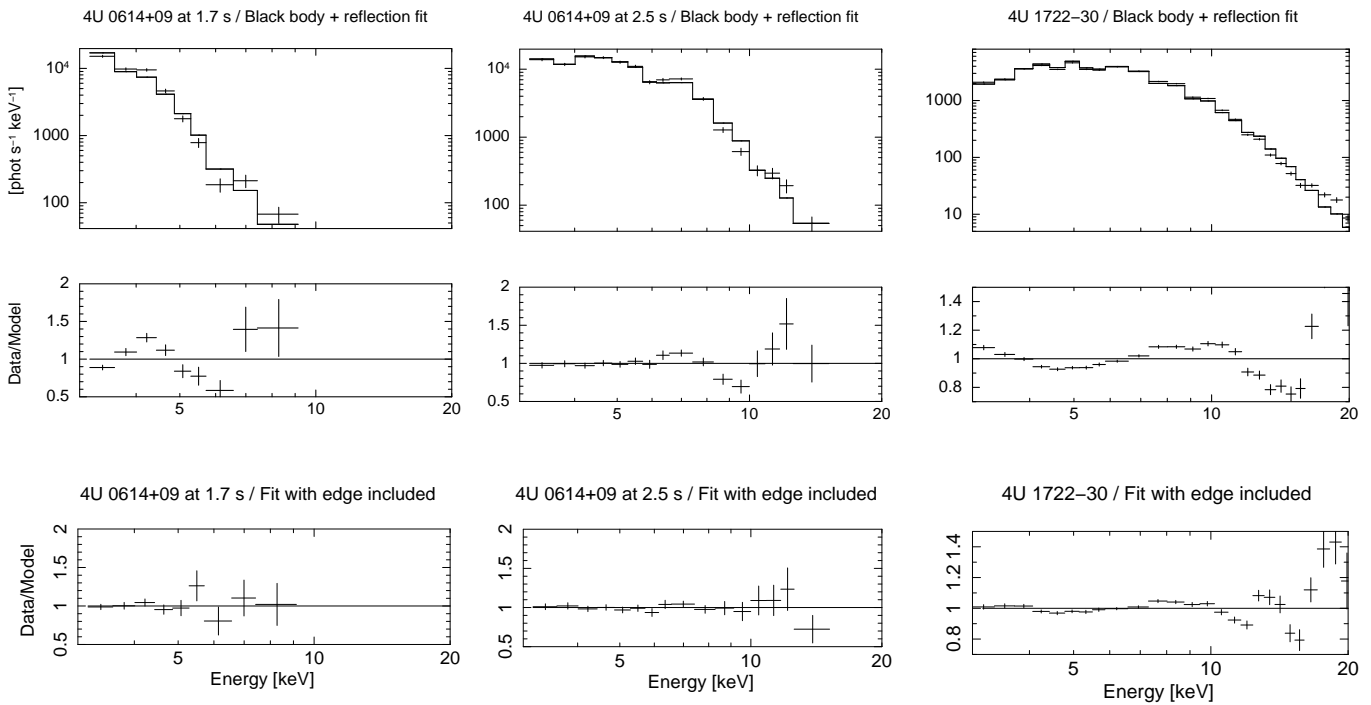

\includegraphics[height=0.675\columnwidth,angle=270]{13952f2a.ps}
\includegraphics[height=0.62\columnwidth,angle=270]{13952f2b.ps}
\hspace{0.2cm}
\includegraphics[height=0.645\columnwidth,angle=270]{13952f2c.ps}

\vspace{8mm}
\hspace{0.1cm}
\includegraphics[height=0.655\columnwidth,angle=270]{13952f2d.ps}
\hspace{0.25cm}
\includegraphics[height=0.60\columnwidth,angle=270]{13952f2e.ps}
\hspace{0.34cm}
\includegraphics[height=0.615\columnwidth,angle=270]{13952f2f.ps}
\caption{Spectral fits to the PCA bursts from 4U 0614+09 and 4U
  1722-30 during time intervals when the absorbed black body plus
  reflection model fails to fit the data (see Fig. \ref{figdev} for
  example time intervals when the black body plus reflection model
  provides a good fit to these bursts). {\em Left plots:} The 4U
  0614+09 spectrum between 1.65 and 1.78 s after burst onset (this is
  the second data point after the superexpansion phase in
  Fig.~\ref{figsp}). {\em Middle plots:} The 4U 0614+09 spectrum
  between 2.53 and 2.65 s after burst onset (ninth data point). {\em
    Right plots:} The 4U 1722-30 spectrum for the 7s long interval
  indicated by the grey bar in Fig.~\ref{figsp}. The upper panels show
  the observed photon flux (crosses) and the fitted black body plus
  reflection model (histogram). The middle panels show the fractional
  deviation of the data points from the black body plus reflection
  model. The lower panels show the deviations after an absorption edge
  is included in the model.
\label{figedge}}
\end{figure*}

The data for the three PCA superexpansion bursts are of higher quality
than the WFC burst data due to the PCA's large photon collecting
area. The quality is especially good because at least three PCUs were
employed during these observations (this is generally true of only
$\approx 5$\% of the PCA's observation time). Nonetheless, the
spectroscopic resolution of the PCA is modest (17\% FWHM at 6 keV) and
only relatively broad spectral features can be resolved.

High-time-resolution spectrally-resolved data are available for the
first 30 s of the burst from 4U 0614+091. Such data are available for
much of the 4U 1722-30 burst, except between 3 and 8 s after burst
onset. The superburst from 4U 1820-30 had good spectral coverage
throughout but only at a modest time resolution of 16 s (the
`Standard-2' data collecting mode). We describe the spectral data and
analysis method in more detail in Appendix \ref{sec:app3bursts}.

\begin{figure*}[t]
\includegraphics[width=0.71\columnwidth,angle=0]{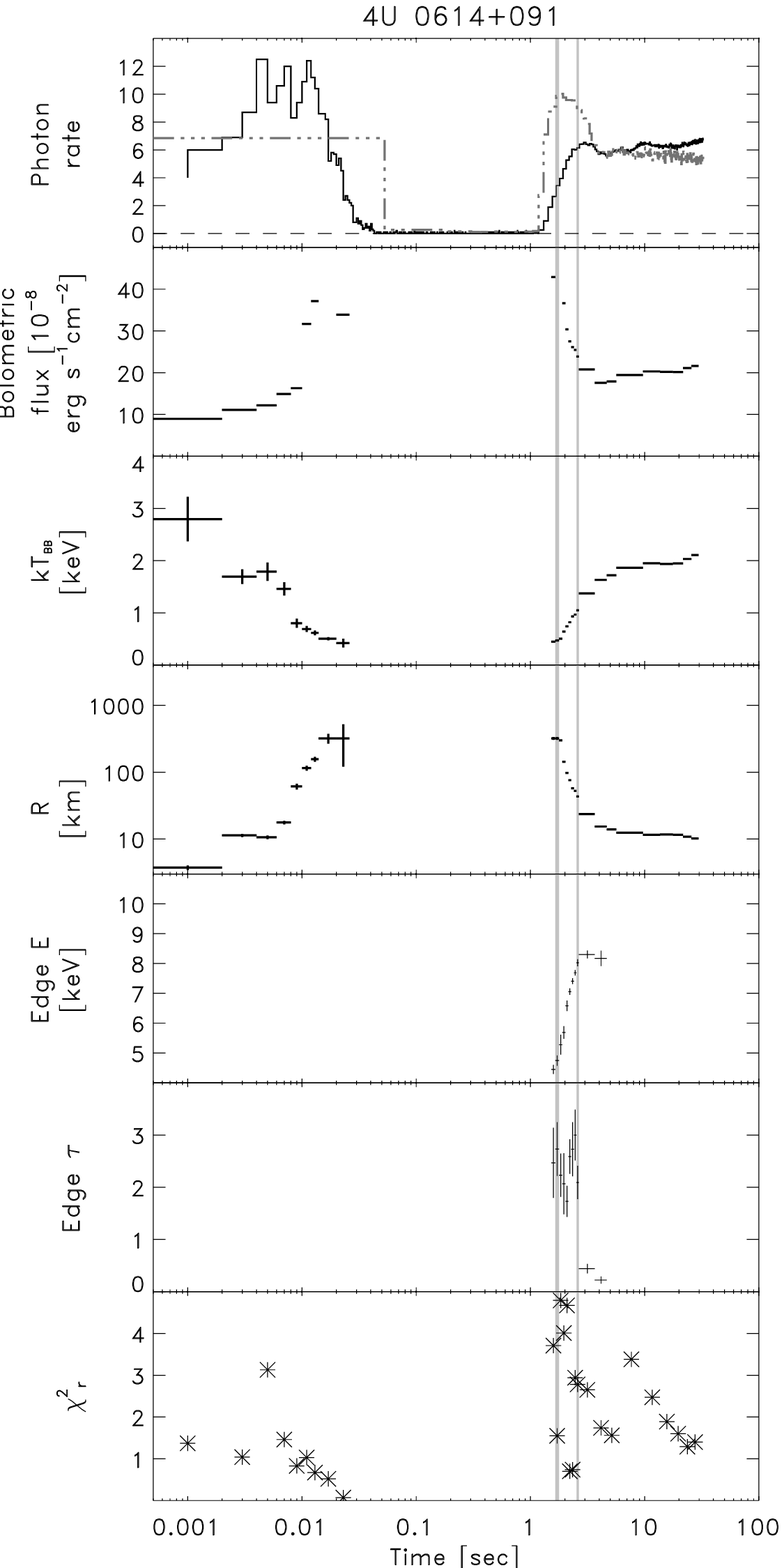}
\includegraphics[width=0.645\columnwidth,angle=0]{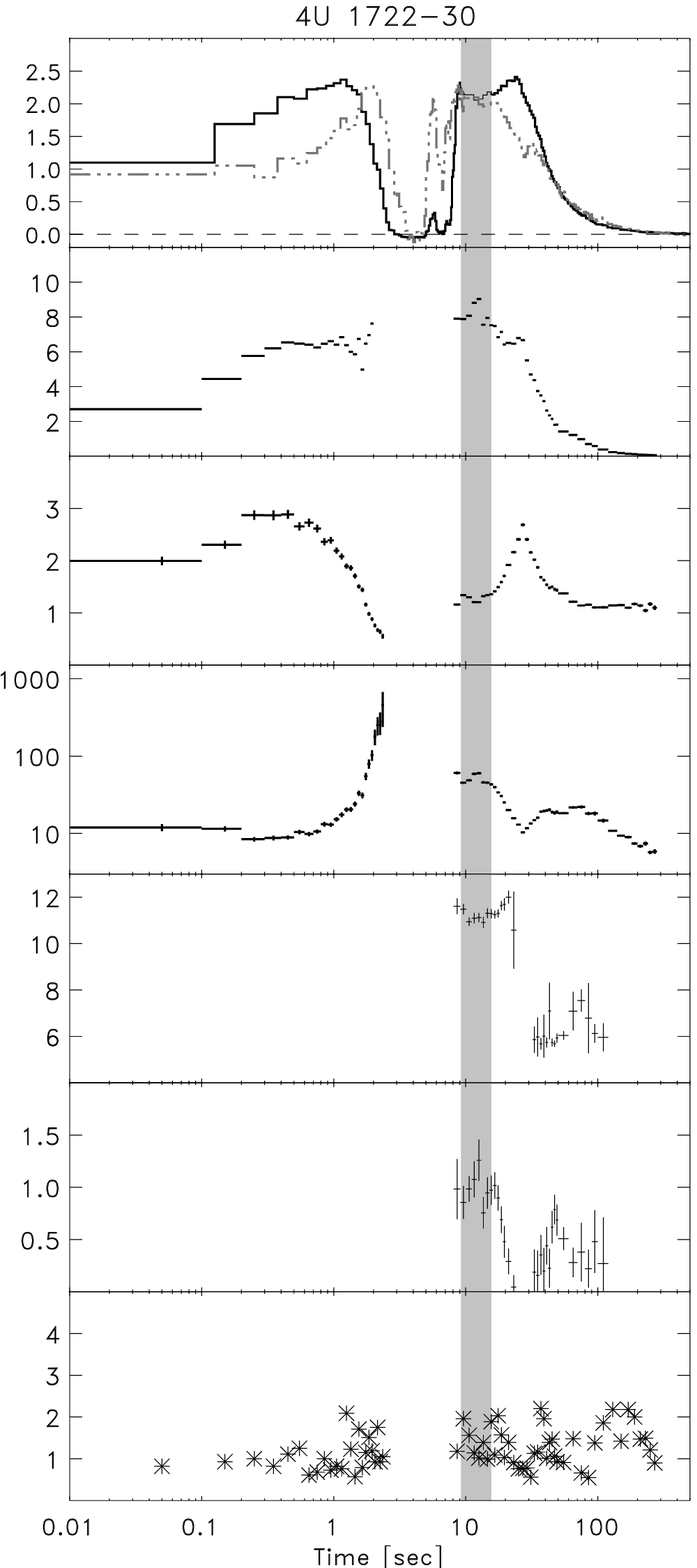}
\includegraphics[width=0.645\columnwidth,angle=0]{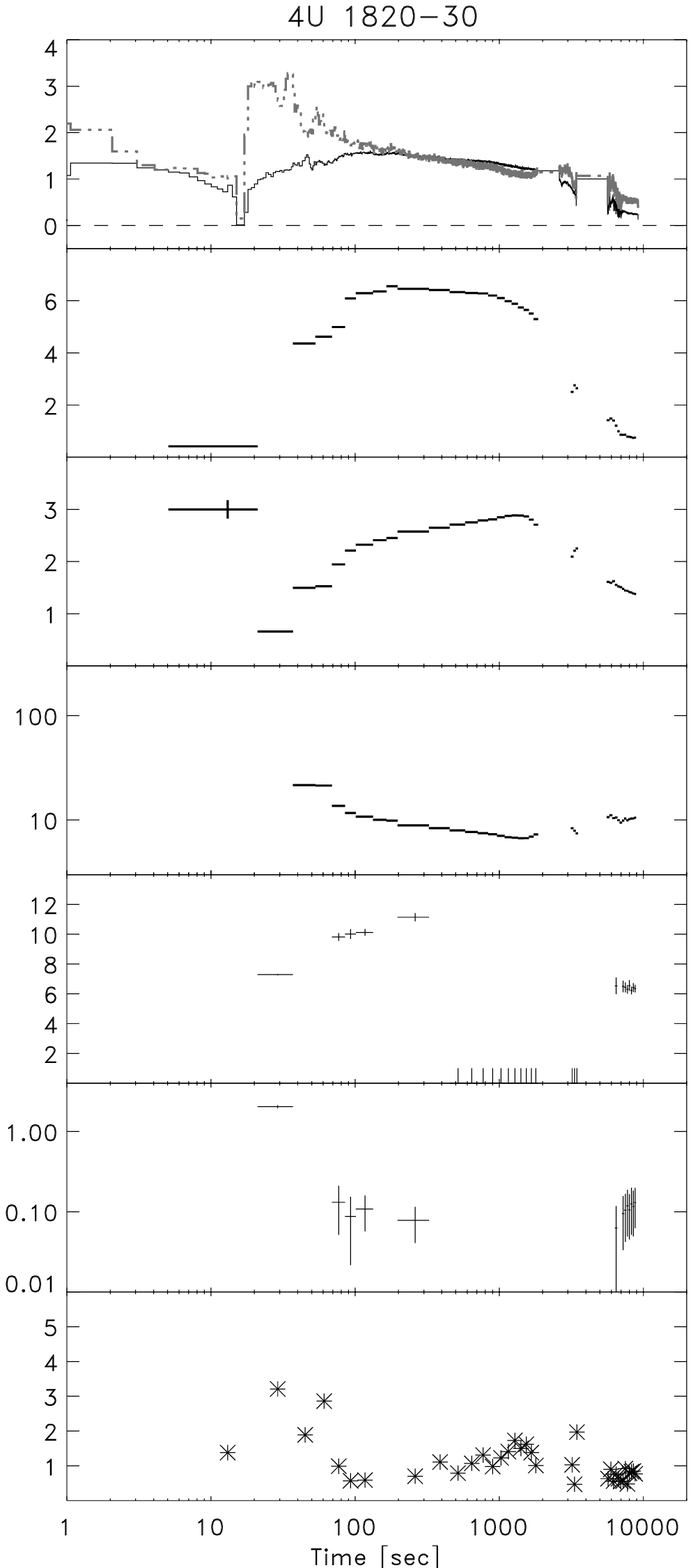}
\caption{Time profiles of various spectral parameters for 3
  superexpansion bursts detected with RXTE-PCA from 4U 0614+09
  \citep[left; see also][]{kuu09}, 4U 1722-30 \citep[center; see
    also][]{mol00} and 4U 1820-30 \citep[right; see
    also][]{stroh02a,bal04}. Time is in seconds since burst onset (see
  Table~\ref{tabobspca}).  {\em Top panels:} total PCA countrate for
  the xenon layers (solid line, in units of 10$^4$~s$^{-1}$) and
  propane layer (dot-dashed grey line, in units of 1,250~s$^{-1}$),
  after subtracting the pre-burst levels. The horizontal dashed lines
  mark a count rate of zero. {\em Second panels:} bolometric flux of
  the fitted black body if no reflection component is included,
  corrected for detector dead time and for absorption. The errors were
  not determined, but they become large when k$T<1$ keV. {\em Third
    panels:} black body temperature in keV. {\em Fourth panels:} black
  body radius assuming isotropic emission. Assumed distances: 3.2 kpc
  \citep[4U 0614+09;][]{kuu09}, 9.5 kpc \citep[4U 1722-30;][]{kuu03}
  and 7.6 kpc \citep[4U 1820-30;][]{kuu03}; {\em Fifth and sixth
    panels:} absorption edge energy $E_{\rm edge}$ in keV and optical
  depth $\tau_{\rm opt}$. {\em Bottom panels:} reduced $\chi^2$ value
  for each fit to the black body + reflection + edge model.  We left
  out the data for 4U 1820-30 between 2000 and 2500 s, because these
  are not understood \citep[see][]{bal04}. The grey vertical bars for
  4U 0614+09 and 4U 1722-30 delimit the time intervals for the spectra
  shown in Fig.~\ref{figedge}. $N_{\rm H}$ was fixed for 4U 0614+09
  and 4U 1820-30 and free for 4U 1722-30 because a fixed value did not
  yield good fits.
\label{figsp}}
\end{figure*}

Typically burst spectra are well fit by an absorbed black body model
\citep{swa77,hof77,stroh06}. Deviations from this model have been
reported for the superexpansion bursts from M15 X-2 \citep{jvp90} and
the PCA superexpansion superburst from 4U 1820-30 (SB02). In the
latter case, SB02 detected a broad emission line between 5.8 and 6.4
keV and an absorption edge at $8-9$ keV at times $>100\textrm{ s}$
after burst onset (at earlier times the blackbody temperature was
evolving too quickly for the Standard-2 accumulations to resolve). To
model these features, SB02 included a Gaussian emission line and an
absorption edge in their spectral model. With these components, they
found acceptable fits with reduced $\chi^2\sim1$; both the line and
edge were strongly required to adequately model the data. SB02 further
noted that features of the Gaussian line---its finite width and an
energy centroid that is often significantly less than 6.4
keV---suggested that the line was produced by reflection from a
relativistic accretion disk. Indeed, excellent fits to the data were
obtained by modeling the line as an Fe fluorescence feature from an
accretion disk and accounting for smearing of the absorption edge due
to disk motions (\citealt{bal04} and \citealt{bal04b}).

\subsection{Absorption edges}
\label{sec:abs_edges_3PCA}

As shown in Appendix \ref{app:model} (see top row of
Fig.~\ref{figdev}), the bursts from 4U 0614+091 and 4U 1722-30 also
show significant deviation from an absorbed black body. The deviations
are similar to those seen in the 4U 1820-30 superburst: a broad
emission line at 6 to $7\textrm{ keV}$ and an edge at $\approx
10\textrm{ keV}$ that varies with time.

Motivated by the success of \cite{bal04} at fitting the deviations in
4U 1820-30 with a model that includes a reflection component, we
attempted to fit the spectra of the 4U 0614+091 and 4U 1722-30 bursts
using the same black body plus reflection model as Ballantyne \&
Strohmayer (provided to us by D. Ballantyne, see
Appendix~\ref{app:model}). Although, as shown in Fig.~\ref{figdev},
the model fits some of the spectra, it fails to fit all the spectra,
particularly spectra just after the end of the superexpansion
phase. Examples of when the black body plus reflection model fail are
shown in the top row of Fig.~\ref{figedge}. These include two spectra
of 4U 0614+091 within a few seconds after the superexpansion and one
spectrum of 4U 1722-30 during the first 7 s after the superexpansion
(as indicated by the grey bar in Fig.~\ref{figsp}). In these three
examples, the edges are deeper or at lower energies than the
reflection model can accommodate.

It is worth noting that during the superexpansion, the flux of both
bursts drops below the pre-burst level (see top panel of
Fig.~\ref{figsp}). This may be because the expanding shell and
super-Eddington flux disrupt the disk and temporarily stop the
accretion. Alternatively, the disk is not disrupted and we just do not
see the accretion flux because the expanding shell surrounds and
obscures the inner disk. Whatever the cause, it is perhaps not
surprising that the disk reflection model fails at times immediately
following the superexpansion.  Furthermore, even when the reflection
model does fit the data, it often yields unreasonable results; for
instance, it requires that the majority of the flux be in the
reflection component rather than the direct black body component. This
results in unreasonable values for the black body luminosity and
radius.

The failure of the reflection model to fit some of the spectral data
from 4U 0614+091 and 4U 1722-30 prompted us to add another component
to the spectral model: an absorption edge stemming from the primary
(non-reflected) spherically symmetric burst emission. \citet{wei06a}
have suggested that such a signature should appear in the spectra
of radius expansion bursts due to the presence of heavy ashes of
nuclear burning in the photosphere (see \S~\ref{sec:MEwinds}). To
model such an edge, we multiply the spectral energy distribution by
$M(E)={\rm exp}[-\tau_{\rm opt}\;(E/E_{\rm edge})^{-3}]$ for photon
energies $E>E_{\rm edge}$ and $M(E)=1$ otherwise\footnote{This is
  multiplicative model {\it edge} in XSPEC.}. The fit parameters are
the optical depth of the line $\tau_{\rm opt}$ and the edge energy
$E_{\rm edge}$. Our full spectral model thus accounts for three
features: an absorbed black body, the reflection of the burst flux
from an inner accretion disk, and an absorption edge formed in the
photosphere.
  
We find that including such an edge significantly improves the
fits\footnote{For some spectra the edge can be satisfactorily fit by
  allowing for unrealistic values of $N_{\rm H}$, the black body
  temperature and the accretion flux. However, for the 4U 0614+091
  burst, the edge immediately after the superexpansion has such a low
  energy (see below) that a good fit cannot be obtained even if one
  allows for unrealistic values of these standard spectral
  parameters.}, as illustrated in the lower row of plots in
Fig.~\ref{figedge}. The evolution of the absorption edge parameters
$\tau_{\rm opt}$ and $E_{\rm edge}$ is shown in Fig.~\ref{figsp} for
the three PCA bursts. The absorption edge from 4U 0614+091 exhibits a
peculiar increase in $E_{\rm edge}$ from $4.63\pm0.05$ keV just after
the superexpansion to $8.5\pm0.1$ keV less than 2 s later (see also
left and middle panels of Fig.~\ref{figedge}). For the next 2 s,
$E_{\rm edge}$ remains nearly constant at 8.5 keV, after which no
significant edge is detected. While $E_{\rm edge}$ is increasing,
$\tau_{\rm opt}$ is large (between 2 and 3), after which it becomes
substantially less than 1.

The fit to the first spectral data point from 4U 1722-30 (8~s after
burst onset) yields an absorption edge with $E_{\rm
  edge}\simeq11\textrm{ keV}$ and $\tau_{\rm opt}\simeq1$. The
parameters remain near these values until the temperature peaks at
23~s after onset, after which the parameters quickly change to $E_{\rm
  edge}\simeq 6-7\textrm{ keV}$ and $\tau_{\rm opt}\simeq0.5$. Since
there is no spectral data immediately following the superexpansion
(from 3 to 8~s), we do not know if the increase in $E_{\rm edge}$ seen
in the spectra from 4U 0614+091 occurs in the case of 4U 1722-30.

The absorption edge from 4U 1820-30 is shallow, but still detected
with significance thanks to the high quality signal provided by the
long duration of the burst. This is particularly clear when co-adding
all the data accumulated over a long time span (e.g., all data between
5000 and 10000 s).

Although we obtain reasonable fits to the data once we include an edge
in the model, the limited spectral resolution of the PCA precludes an
unambiguous identification of these features as absorption
edges. Higher spectral resolution observations of superexpansion
bursts, for instance with XMM-Newton or Chandra, would be extremely
useful.

\subsection{Other spectral features of the 3 PCA bursts}
\label{sec:other_PCA_features}

The evolution of some of the other spectral parameters is shown in
Fig.~\ref{figsp}. We do not show the results for the reflection
parameters as these are not of immediate interest. The top panel shows
the light curve for both the xenon and propane detector layers. The
most conspicuous feature is the significant dip in the light curve
during the superexpansion phase; the flux drops below the pre-burst
level signifying that the large burst flux temporarily disrupts or
obscures the accretion on to the NS. The propane layer is most
sensitive at 1.5 to 2 keV and the signal from this layer is strongest
when the black body temperature $kT\approx1\textrm{ keV}$ immediately
following the superexpansion. The light curve from 4U 1722-30 has a 2
s long spike near the end of the superexpansion phase that is
especially pronounced in the propane layer. It may be due to a brief
contraction and re-expansion of the photosphere, although interpreting
the spike is difficult because there is no spectral data available
during this time interval. There are indications of similar spikes in
the 4U 1722-30 bursts detected with the WFC (Fig.~\ref{figtz2wfc}). A
similar feature has been reported in a burst from M15 X-2
\citep{jvp90}.

The 3 PCA bursts show a phase of moderate expansion after the
superexpansion. During the moderate expansion phase, the black body
temperature increases and the radius decreases. We define the end of
the moderate expansion phase to be the time at which the temperature
peaks, which is attributed to the photosphere `touching-down' on the
NS surface \cite[e.g.,][]{lew93,kuu03,gal08,gal08b}. This is nicely
illustrated in Fig.~\ref{figsp} in the bursts from 4U 1722-30 and 4U
1820-30, where touch down occurs 23 and 1400 s after burst onset,
respectively. The moment of touch down in 4U 0614+09 was not covered by
the PCA, but was observed with FREGATE on HETE-II to be at 90 s
\citep{kuu09}.

In the bursts from 4U 0614+09 and 4U 1722-30, we find that at the
start of the superexpansion phase, when the X-ray flux begins to drop,
the photosphere is apparently moving out at velocities $v_{\rm ph}
\approx 3\times10^4\textrm{ km s}^{-1}$ and $v_{\rm ph} \approx
10^3\textrm{ km s}^{-1}$, respectively (left and middle panels of
Fig. \ref{figsp}; this estimate is simply the change in apparent
radius divided by the time over which that change occurred). It is
difficult to determine whether these measurements reflect the true
velocities of the photosphere due to uncertainties, e.g., in the black
body color correction \citep{London:86,Pavlov:91}. Calculations by
\cite{Madej:04} show that the observed color temperature of a black
body fit may differ by a factor of 1.3 to 1.4 from the effective
temperature for a 1 keV black body  and about 1.6 for a 3
keV black body. Consequently, inferred radii would need to be
corrected by a factor of 1.7 to 2.6. In addition, the PCA calibrated
data do not cover photon energies below 3 keV and, thus, the peak of
the thermal spectrum when the expansion is large. We nonetheless
consider the expansion velocities to be reliable to within a factor of
a few.

The radii during the moderate expansion phase of the 4U 1820-30
superburst are unreasonably small (i.e., $<10\textrm{ km}$). It cannot
be attributed to an underestimated value of the distance as this is
well measured both optically \citep{hea00} and from comparing the
burst peak flux to the Eddington limit of a hydrogen-poor photosphere
\citep[e.g.,][]{zdia07,gal08}. A color correction would, however,
increase the measured radii to acceptable values \citep[see
  also][]{stroh02a}. The superburst also behaves somewhat unusually
between 2000 and 2500~s after burst onset. The behavior is unexplained
and the data in this time frame is excluded in both our analysis and
that of \citet{bal04}.

\section{Spectroscopic features of the other bursts}
\label{sec:spectra_nonPCA}

When there is sufficient sensitivity, other superexpansion bursts show
some or all of the features seen in the 3 PCA bursts. The edge is
difficult to detect in bursts measured with instruments other than the
PCA, although the burst from M15 X-2 measured with Ginga showed
indications of edges in the spectrum (see Fig. 6 in
\citealt{jvp90}). The one feature that is {\em always} seen is a phase
of moderate expansion following the superexpansion.  The spike in 
  low-energy flux a few seconds after superexpansion is not always
seen (e.g., it is not seen in the PCA bursts from 4U 0614+09 and 4U
1820-30).

The burst from 4U 0614+09 has a very fast rise and the precursor is
merely 0.05~s long (Fig.~\ref{figsp}).  For such short durations,
smaller detectors might fail to catch the precursor and instead first
see the burst during the end of the superexpansion phase, when the
photospheric radius is decreasing. An example of this might be the
long burst detected from 2S 0918-549 with the BeppoSAX WFCs
\citep{zan05a}. This burst lacks a precursor and yet spectroscopic
analysis of the rising phase reveals a black body radius that is
rapidly decreasing, as if the burst was preceded by a superexpansion
phase. The very long burst from SLX 1737-282 \citep[$\tau_{\rm
    decay}=659\pm32$~s, see also][]{zan02,fal08} also lacks a
precursor despite having a radius that decreases during the rise,
although the decrease is not as extreme as in 2S 0918-549.

There are a few bursts that also show two expansion phases, but whose
expansion factor in the first phase is too small (between 5 and 10) to
qualify as a superexpansion. Their X-ray signal is not completely
lost, but their light curve still shows a deep dip. This is true of
all ordinary X-ray bursts detected from 4U 1820-30 with the PCA
\citep{gal08}, a second PCA burst from 4U 1722-30 detected at MJD
54526.679 (Galloway, priv. comm.), a burst from M15 X-2 \citep{sma01},
and a few bursts from the non-UCXBs 4U 1705-44, KS 1731-260, SAX
J1747.0-2853, and SAX J1808.4-3658.

\section{Measurement of expansion durations}
\label{sec:texpansion}

\begin{figure}[t]
\includegraphics[height=\columnwidth,angle=0]{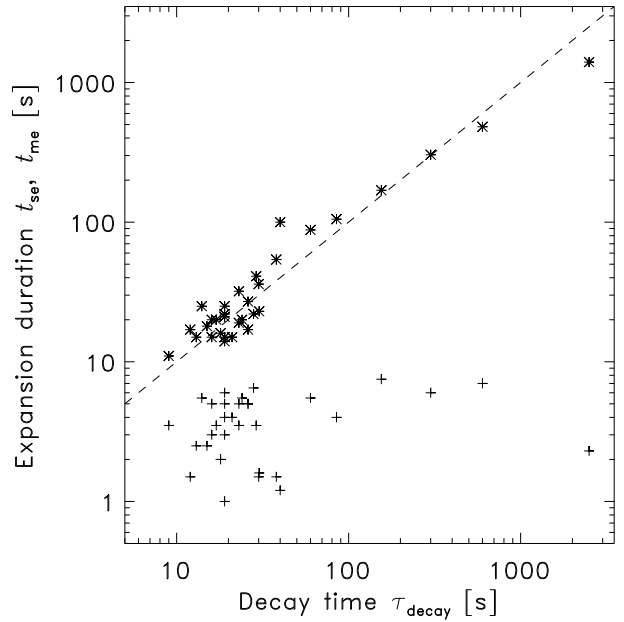}
\caption{Duration of radius expansion phases versus decay time
  $\tau_{\rm decay}$ for each superexpansion burst. For data, see
  Table~\ref{tab1}. Crosses mark the duration of a burst's
  superexpansion phase ($t_{\rm se}$) and asterisks mark the duration
  of its moderate expansion phase ($t_{\rm me}$). The dashed line
  marks the relationships $t_{\rm se},t_{\rm me}=\tau_{\rm
    decay}$. \label{figdiagram}}
\end{figure}

We define the duration of the superexpansion phase, $t_{\rm se}$, as
the time between the end of the precursor and the start of the main
burst. In practice (allowing for statistical uncertainties), this is
the time over which the flux is less than about $10\%$ of its peak
value. We define the duration of the moderate expansion phase, $t_{\rm
  me}$, as the time between the end of the superexpansion and the time
of peak temperature or spectral hardness, which is usually considered
to be the moment when the expanded photosphere `touches down' on the
NS (see \S~\ref{sec:pcaspec}). The decay time $\tau_{\rm decay}$ is
determined by fitting an exponential function to the full-bandpass
light curve after touch down and measuring the e-folding decay time.
Since all the detectors are of a similar type (i.e., a xenon-filled
proportional counter), the systematic errors introduced by using {\it
  observed} photon rates instead of {\it derived} calibrated energy
fluxes are negligible compared to statistical errors.

Table~\ref{tab1} gives the duration measurements for the 32
superexpansion bursts.  The durations $t_{\rm se}$ and $t_{\rm me}$
are plotted as a function of $\tau_{\rm decay}$ in
Fig.~\ref{figdiagram}.  The diagram reveals two important features:
first, $t_{\rm me}$ is proportional to $\tau_{\rm decay}$ (they are
approximately equal). This is not surprising; the longer the burst,
the longer the flux remains near its peak value. Since the peaks are
super-Eddington in these bursts, the duration of the super-Eddington
phase should be proportional to the decay time (see
Appendix~\ref{app:unlikely_tse_explanations}). Second, $t_{\rm se}$ is
independent of $\tau_{\rm decay}$; in all bursts the superexpansion
lasts a few seconds.  We discuss possible explanations for the lack of
correlation between $t_{\rm se}$ and $\tau_{\rm decay}$ in
\S~\ref{sec:tse_discussion}.

\section{Discussion}
\label{sec:discussion}

\subsection{Association between superexpansion bursts and UCXBs}
\label{sec:ucxb}

All the superexpansion bursts that we find, including those in the
literature, are from (candidate) UCXBs \citep[see][]{zan07}. A
characteristic of UCXBs is that they are hydrogen deficient
\citep[e.g.,][]{nel08}. As a result, X-ray bursts from UCXBs are most
likely fueled entirely by helium and/or heavier elements\footnote{It
  is unclear whether free-falling $\alpha$ particles, in the absence
  of protons, will spall heavier accreted nuclei already present in
  the NS atmosphere to such a degree that significant amounts of
  protons are produced
  \citep[cf.,][]{bil92,Bildsten:03}.}. Furthermore, since there is no
stable CNO burning in-between bursts, the freshly accreted NS envelope
in an UCXB is colder than that of a hydrogen-rich system accreting at
the same rate. Bursts from UCXBs therefore tend to occur at greater
depth, where the thermal time is longer. This gives rise to so-called
`intermediate-duration bursts' with durations of several tens of
minutes \citep{zan05a,cum06}. Because of the large and rapid energy
release, the peak luminosity of helium-rich bursts (of both the short-
and intermediate-duration variety) often exceeds the Eddington limit,
resulting in phases of photospheric radius expansion
\citep{Paczynski:83, wei06a, cum06}.

\subsection{Evidence for heavy-element ashes in photosphere}
\label{sec:MEwinds}

The convective region that forms during an X-ray burst is well-mixed
with the freshly synthesized ashes of nuclear burning. \citet{wei06a}
solve for the evolution of the vertical extent of the convective
region in PRE bursts with strong radiative winds. They find that the
convective region extends outwards to sufficiently shallow depths that
the base of the wind lies well within the ashes of burning. As a
result, some ashes are ejected in the wind and exposed at the
photosphere. The column density of ejected ashes, which consist
primarily of heavy elements (near the Fe peak; see below), is expected
to be especially large for the energetic, helium-rich bursts that
exhibit superexpansion. The spectral signature of the ashes is
therefore expected to be detectable at even modest spectral resolution
(\citealt{wei06a}; see also \citealt{vanParadijs:82} and
\citealt{Foster:87}).

We now discuss two features in the data that suggest the presence of
heavy-element ashes in the photosphere: absorption edges at the end of
the superexpansion phase (\S~\ref{sec:disedges}) and
smaller-than-expected radii during the moderate expansion phase
(\S~\ref{sec:smallradii}).

\subsubsection{Absorption edges}
\label{sec:disedges}

Which elements might be responsible for the observed absorption edges?
We first consider the early-time edge in the spectrum of 4U 0614+09,
which increases from an initial energy of $E_{\rm edge}=4.6 \textrm{
  keV}$ to 8.5 keV within $\simeq 2\textrm{ seconds}$. Since the
photosphere is at large radii ($r_{\rm ph} \ga 30\textrm{ km}$) during
this time, the gravitational redshift should be negligible ($z\la
0.1$). Although $r_{\rm ph}$ is decreasing with time, suggesting that
material is moving away from the observer and back on to the NS, the
rate of decrease appears much too slow ($\sim 100\textrm{ km s}^{-1}$)
to produce any appreciable doppler redshift.\footnote{If instead it
  was a radial free-fall, the redshift could be significant even at
  these large radii. The ratio of observed to emitted frequency would
  be $\nu_{\rm obs}/\nu_{\rm em} =1-\beta$ where $\beta=v/c=(2
  GM/c^2r)^{1/2}$ is the free fall velocity. Thus, for
  $r\simeq30\textrm{ km}$ have $\nu_{\rm obs}\simeq0.6 \nu_{\rm em}$.}
This suggests that the variations in $E_{\rm edge}$ are not due to
variations in redshift but rather are intrinsic to the source.  One
possibility is that the shift in $E_{\rm edge}$ is due to variations
in which elements and ionization states dominate the absorption as the
photosphere contracts and $T_{\rm eff}$ increases.  If correct,
bound-bound transitions should also be a source of absorption. The
limited spectral resolution of the PCA makes it difficult to detect
narrow X-ray absorption lines, however (such lines have never been
confirmed in PCA spectra as far as we know).

We now consider the late-time 8.5 keV edge in 4U 0614+09. The measured
photospheric radii at this time ($r_{\rm ph}\approx 10\textrm{ km}$)
imply that the photosphere is close to the NS surface and, thus, that
the edge is gravitationally redshifted. For typical NS parameters
(e.g., $M=1.4M_\odot$, $R=10\textrm{ km}$) $z\approx0.3$ at the
surface and the intrinsic energy of the $8.5 \textrm{ keV}$ edge would
be $11\textrm{ keV}$. This is close to both the hydrogen-like and
helium-like edges of Ni ($10.8\textrm{ keV}$ and $10.3\textrm{ keV}$,
respectively). That the edge may be due to Ni is supported by
calculations which show that at the large ignition depth\footnote{We
  choose to express depth in terms of the mass contained in a column
  of unit cross-sectional area.}  implied by the intermediate duration
  of the burst from 4U 0614+09 ($y_{\rm base}\sim10^{10}\textrm{ g
    cm}^{-2}$; \citealt{kuu09}), Ni is the dominant product of
  constant pressure helium burning \citep{Hashimoto:83}.

The 4U 1722-30 moderate expansion phase edge at $E_{\rm edge}\approx
11\textrm{ keV}$ is also close to the unshifted hydrogen-like Ni edge,
consistent with the large measured photospheric radius during this
phase. After the moderate expansion phase, when the photosphere is at
the NS surface, $E_{\rm edge}\approx 7\pm1 \textrm{ keV}$, consistent
with a gravitationally redshifted hydrogen-like Ni edge.
   
We can also use the measured optical depth of the edge, $\tau_{\rm
  opt}$, to demonstrate that not only must the absorption be due to a
heavy element such as Ni, but that its abundance must be much higher
than the solar value. We do so by assuming that $\tau_{\rm opt} =
\tau_{\rm bf}(E_e) \equiv N \sigma_{\rm bf}(E_e)$, where $N$ is the
hydrogenic column density of the element above the photosphere,
$\sigma_{\rm bf}(E) \simeq 6.3\times10^{-18}(E_{\rm edge}/E)^3
Z^{-2}\textrm{ cm}^2$ is the bound-free cross-section, and we take
$E_{\rm edge}= 13.6Z^2\textrm{ eV}$ (see, e.g.,
\citealt{Bildsten:03}). We approximate the hydrogenic column density
as $N(T, \rho, Z)= f \zeta_H N_e$, where $f$ is the fraction of atoms
belonging to that element (the `number fraction'), $\zeta_H $ is the
fraction in the hydrogen-like state at a given temperature and density
from Saha equilibrium (we assume an element is either fully ionized or
hydrogen-like), $N_e \approx \sigma_{\rm Th}^{-1}$ is the electron
number column density, and $\sigma_{\rm Th}$ is the Thomson
cross-section. Such an approximation should give a reasonable estimate
of $N$ (see, e.g., \citealt{wei06a}).

\begin{figure}
\includegraphics[bb= 45 180 0 700, height=10.5cm,angle=0]{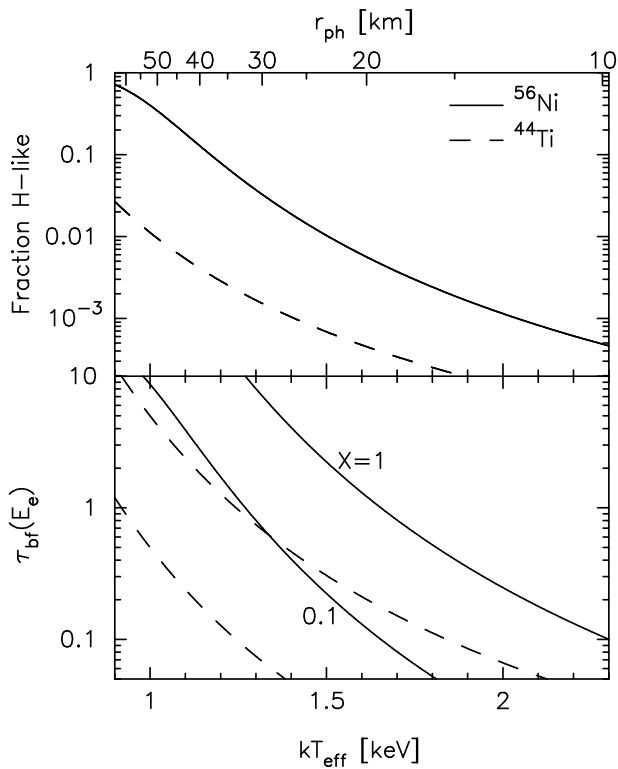}
\caption{Fraction of  $^{44}$Ti and $^{56}$Ni in a hydrogenic state
  $\zeta_H$ ({\em top panel}) and the optical depth of the edge
  $\tau_{\rm bf}(E_e) \equiv N \sigma_{\rm bf}(E_e)$ ({\em bottom
    panel}) as a function of effective temperature ({\em bottom axis})
  and blackbody radius $r_{\rm ph}\equiv (L_{\rm Edd} / 4\pi \sigma
  T_{\rm eff}^4)^{1/2}$ ({\em top axis}) assuming a density of $\rho
  =0.01 \textrm{ g cm}^{-3}$. Results are shown for mass fractions of
  $X=0.1$ and 1.\label{figopacity}}
\end{figure}

In the top panel of Fig.~\ref{figopacity} we show $\zeta_H$ as a
function of effective temperature $kT_{\rm eff}$ and $r_{\rm
  ph}=(L_{\rm Edd} / 4\pi \sigma T_{\rm eff}^4)^{1/2}$ for $^{44}$Ti
and $^{56}$Ni at a density\footnote{In the wind models, $d\ln
  \rho/d\ln r \ga -3$ and $\rho\approx 1\textrm{ g cm}^{-3}$ at $r=R$
  (see \citealt{pac86}); since the observed expansion factors are
  $\approx 3-5$ during the moderate expansion phase, we expect the
  photospheric density to be $\approx 10^{-2} \textrm{ g
    cm}^{-3}$. For $\zeta_H\la0.1$, the optical depth $\tau_{\rm bf}$
  scales almost linearly with $\rho$.}  of $\rho=0.01\textrm{ g
  cm}^{-3}$.  In the bottom panel we show the implied optical depth of
the edge $\tau_{\rm bf}(E_e) \equiv N \sigma_{\rm bf}(E_e)$ for mass
fractions of $X=0.1$ and $1$.  As the figure shows, a mass fraction
$X\ga 0.1$ is required in order to obtain an optical depth as large as
the measured values $\tau_{\rm opt}\ga 1$ at the observed
temperatures. Furthermore, elements $Z\la22$ cannot produce
sufficiently large optical depths even for $X=1$. A Ni (Fe) mass
fraction of $X\sim0.1$ is $\sim 1000$ ($100$) times larger than the
solar abundance.

Rotational broadening is expected to be small compared to the PCA
spectral resolution and the width of an absorption edge. The NS in 4U
0614+09 has a spin frequency of $\nu_{\rm spin}=415$ Hz
\citep{stroh08}, resulting in a maximum relative Doppler shift of
$v/c=2\pi\nu_{\rm spin}R_{\rm NS}/c \simeq 0.1$ and a FWHM of the
broadening profile approximately a factor 2 lower (n.b., viewing angle
effects and the smaller rotation velocity of an expanded photosphere
tend to further reduce the broadening). The other two bursters have no
measured spin frequencies, but the highest confirmed spin frequency
ever measured for a burster is only 50\% larger
\citep[620~Hz;][]{muno02}.

\subsubsection{Small radii during moderate expansion phase}
\label{sec:smallradii}

Steady state models of radiative winds from NSs predict photospheric
radii of $100-1000 \textrm{ km}$ for mass outflow rates of
$\dot{M}\approx10^{17}-10^{19}\textrm{ g s}^{-1}$ \citep{pac86,
  Joss:87, Nobili:94}. The measured photospheric radii during the
moderate expansion are much smaller than this ($r_{\rm ph} \approx
30-50 \textrm{ km}$).  The models were not calculated for lower
$\dot{M}$ for numerical reasons and perhaps the observed radii are
small because the outflow rates are $\dot{M} < 10^{17} \textrm{ g
  s}^{-1}$. This seems unlikely, however.  From energy conservation,
\citet{pac86} showed that the radiatively-driven mass outflow rate
from a neutron star is given by $\dot{M} \simeq \alpha \dot{M}_0$,
where $\alpha\equiv (L_{\rm base} - L_{\rm Edd}) / L_{\rm Edd}$ is the
excess luminosity at the base of the ignition layer relative to
Eddington and $\dot{M}_0 = 1.7\times10^{18}\textrm{ g s}^{-1}$ is a
constant that depends only on the neutron star mass and radius (taken
to be $M=1.4M_\odot$ and $R=10\textrm{ km}$). Such a low $\dot{M}$
(i.e., $\alpha \la 0.05$) would therefore seem to require fine-tuning
as it would imply that all the bursts are super-Eddington by only the
slightest amount.

An alternative possibility is that the outflow structure is truncated
at a radius where absorption becomes sufficiently more important than
electron scattering; material above this radius has a lower effective
Eddington limit and is blown away. The wind models either ignored
absorption \citep{ebi83, pac86, Nobili:94} or assumed the wind
consists of solar abundances \citep{Joss:87}. The presence of heavy
element ashes in the wind can, however, significantly modify the
structure of the outflow.

To see why, note that at the temperatures observed during the moderate
expansion phase ($kT_{\rm eff} \simeq 1 - 2 \textrm{ keV}$), the
fraction of Ni in the hydrogen-like state is $0.001-0.4$ from Saha
equilibrium at a density $\rho=0.01\textrm{ g cm}^{-3}$ (this assumes
the Ni is either fully ionized or hydrogen-like; Ni is the dominant
product of burning at these depths, see \S~\ref{sec:disedges}). The
radiative force on hydrogenic Ni in the ground state is, in the
optically thin limit (i.e., ignoring line blanketing), $F_{\rm rad} =
\pi h e^2 f_{1\rightarrow 2} F(E)/ m_e c^2$ \citep{Bildsten:03,
  Michaud:70}, where the flux per unit energy at the Ly$\alpha$
transition $F(E)$ is nearly constant across the line ($E=8.0\textrm{
  keV}$) and $f_{1\rightarrow2}=0.42$ is the oscillator strength for
hydrogenic Ni \citep{Fuhr:81}. At the observed temperatures, the
radiative acceleration on Ni is then $a_{\rm rad}\approx
10^{15}-10^{16} \textrm{ cm s}^{-2}$ assuming a blackbody
spectrum\footnote{The spectra of type I bursts are somewhat harder
  than a blackbody at the observed $T_{\rm eff}$ \citep{London:86,
    Pavlov:91, Madej:04, maj05}. The radiative force acting on
  hydrogenic ions may therefore be larger than our estimates, although
  the ionization fraction will also be smaller.}, a factor of $\ga
100$ larger than the gravitational acceleration at the observed
photospheric radii $r_{\rm ph}\approx30-50\textrm{ km}$. The force
acting on the bound electrons is therefore $\ga 100$ times larger than
that acting on the free electrons and line driving will become
dynamically significant once the abundance ratio of bound to free
electrons exceeds $\sim 1/100$ (see, e.g., \citealt{Gayley:95}). This
suggests that once material reaches $\approx 50\textrm{ km}$ and a
significant fraction of ions become hydrogenic, line-driving will
accelerate the flow to velocities much higher than the $v\approx
10^3\textrm{ km s}^{-1}$ found in the aforementioned wind models.

Since $\rho\propto v^{-1}$ in a steady state wind, the structure of
the outflow above $\approx 50\textrm{ km}$ may be much more tenuous
than the wind models suggest. If line-driving accelerates the material
to $\sim 10^4-10^5 \textrm{ km s}^{-1}$, the column density of
overlying material will be a factor of $\sim 10-100$ smaller than that
found in the wind models. The radius of the electron scattering
photosphere may therefore be smaller by the same factor, in better
agreement with the observed $r_{\rm ph}$. One caveat is that it is not
clear whether lines or electron scattering will be the dominant source
of opacity; a wind calculation that accounts for the absorption
opacity from heavy elements is therefore needed in order to reliably
estimate the location of the photosphere. While the electron
scattering opacity dominates the free-free and bound-free opacity for
the measured temperatures and densities of the photosphere (even if
the composition is primarily heavy elements; see, e.g.,
\citealt{Cox:68}), the bound-bound opacity may be important.

\subsection{Superexpansion duration}
\label{sec:tse_discussion}

If superexpansion is due to a super-Eddington flux driving the
photosphere to large radii, as the time profiles suggest, why is the
superexpansion duration $t_{\rm se}$ always short (a few seconds) and
independent of the decay time $\tau_{\rm decay} \sim 10-1000 \textrm{
  s}$ (see \S~\ref{sec:texpansion})? Since the longer a burst, the
longer the flux remains near its peak value, one might have expected
$t_{\rm se}$ to increase with increasing $\tau_{\rm decay}$, as is the
case with the moderate expansion duration $t_{\rm me}$.

We now describe a possible explanation for this lack of correlation
between $t_{\rm se}$ and $\tau_{\rm decay}$. In Appendix
\ref{app:unlikely_tse_explanations} we show why the lack of
correlation is unlikely the result of either a brief (second-long)
phase of highly super-Eddington flux or hydrodynamic burning.

X-ray burst wind models have all assumed a steady
state outflow \citep{ebi83, pac86, Joss:87, Nobili:94} and do not
therefore describe the time-dependent response of the overlying layers
in the very first moments of radiative driving. These models do,
however, show that it takes $\approx 1 \textrm{ s}$ for
time-independent conditions to be established (see $t_{\rm char}$ in
table 1 of \citealt{Joss:87}). Given that $t_{\rm se} \approx 1
\textrm{ s}$, this suggests that the superexpansion may be the result
of a transient stage in the development of the wind.

Analyzing this possibility requires a time-dependent wind calculation
and is beyond the scope of this paper. It is nonetheless worth noting
that the superexpansion is consistent with the ejection of a
geometrically thin shell of material of initial column depth $y_{\rm
  ej} \gg y_{\rm ph}$, where $y_{\rm ph}\simeq 1 \textrm{ g cm}^{-2}$
is the depth of the photosphere on the NS surface. This shell, which
may correspond to the layer of material above where the flux first
becomes super-Eddington\footnote{During the burst rise, the flux first
  exceeds the Eddington limit at some depth deep below the
  photosphere. This is because the electron scattering opacity
  increases towards the surface due to the temperature-dependent
  Klein-Nishina corrections \citep{Paczynski:83}. A given flux can
  therefore be sub-Eddington in the deeper layers and at a certain
  point become super-Eddington as it diffuses upward into lower
  density, cooler material.}, will become transparent when it reaches
an expansion factor $r/R \sim (y_{\rm ej}/y_{\rm ph})^{1/2}$. Once it
becomes transparent---at a time $t_{\rm se}$ according to this
picture---the observer suddenly sees the underlying neutron star
again. Since the wind has had enough time to reach a steady state by
this time, the photosphere is located at the moderate expansion value
$r_{\rm ph} \approx 30-50 \textrm{ km}$.

The nuclear energy release from He burning ($\approx 1 \textrm{ MeV
  nucleon}^{-1}$) is approximately $1\%$ of the gravitational binding
energy at the NS surface, and therefore at most $1\%$ of the accreted
mass can be ejected. This implies $y_{\rm ej} \la 0.01 y_{\rm
  base}$. If we assume the velocity of the ejected material is roughly
constant with radius, the initial depth of the ejected column is
$y_{\rm ej} = y_{\rm ph} (v_{\rm ph} t_{\rm se} / R)^2$ and we have
the constraint $y_{\rm base} \ga 10^8 (v_{\rm ph} t_{\rm se} / 10^4
\textrm{ km})^2 \textrm{ g cm}^{-2}$. Since even short duration X-ray
bursts have ignition depths $y_{\rm base} > 10^8 \textrm{ g cm}^{-2}$,
this constraint is easily satisfied in the case of 4U 1722-30 ($v_{\rm
  ph} \approx 10^3 \textrm{ km s}^{-1}$ and $t_{\rm se} = 5 \textrm{
  s}$; see \S~\ref{sec:other_PCA_features} and Table~\ref{tab1}). In
the case of 4U 0614+09, the only other superexpansion burst for which
there is an estimate of $v_{\rm ph}$, we get the strong constraint
$y_{\rm base} \ga 10^9 \textrm{ g cm}^{-2}$ for $v_{\rm ph} \approx
3\times10^4 \textrm{ km s}^{-1}$ and $t_{\rm se} = 1.2 \textrm{
  s}$. This burst is an intermediate-duration burst and the ignition
depth inferred from the cooling time is indeed consistent with $y_{\rm
  base}\sim10^{10} \textrm{ g cm}^{-2}$ \citep{kuu09}.

\section{Summary and conclusions}
\label{sec:conclusions}

The volume and quality of data on superexpansion bursts has increased
substantially since the first detailed study nearly 20 years ago
\citep{jvp90}. We found 28 new superexpansion bursts in the burst
catalogs and literature, several of which were detected with much
higher temporal and spectral resolution than the 4 original
superexpansion bursts.  In studying this larger sample, we found that
superexpansion bursts have the following features: 
\begin{enumerate}
\item At least 31 of the 32 superexpansion bursts are from
  (candidate) ultracompact X-ray binaries.
\item Three PCA bursts show significant spectral deviations from
  the absorbed black body model that typically describe X-ray burst
  spectra. The spectral deviations can be explained as absorption
  edges, with edge energies and depths that vary with time. 
\item Superexpansion is seen in short duration bursts,
  intermediate-duration bursts, and superbursts and the superexpansion
  phase is always followed by a moderate expansion phase. 
\item The duration of the superexpansion is always a few seconds,
  independent of burst duration, while the duration of the moderate
  expansion phase is proportional to burst duration. 
\item The photospheric radius during the moderate expansion phase
  is significantly smaller than that predicted by models of
  radiation-driven X-ray burst winds.
\end{enumerate}

We showed that the absorption edges and small photospheric radii of
the moderate expansion phases may indicate the presence of
heavy-element ashes in the wind. In particular, the edge energies and
optical depths appear consistent with a redshifted hydrogen-like or
helium-like Ni edge and high Ni mass fractions ($X\ga0.1$); Ni is
expected to be the dominant product of burning at the large ignition
depths inferred from the duration of these bursts.  While we do not
have a good explanation for why the edge energies seem to shift over
the course of a few seconds, we speculate that it may be due to
variations in which elements and ionization states dominate as $r_{\rm
  ph}$ decreases and the temperature increases.

The brevity of the superexpansion and its lack of correlation with
burst duration may be the result of a transient stage in the wind's
development, perhaps during which a shell of material is ejected to
large radii by the sudden onset of super-Eddington flux deep below the
photosphere.  A time-dependent wind calculation using realistic
compositions and opacities is  needed in order to assess
the viability of this proposed explanation.

The spectral deviations seen in the PCA bursts suggest that the
spectra of superexpansion bursts have the potential to provide great
insight into burst and NS physics. High spectral resolution
observations with sub-2 keV energy coverage, as can be achieved with
{\it Chandra} or {\it XMM-Newton}, are clearly needed. The shortest
burst recurrence times for X-ray sources that exhibit almost
exclusively superexpansion bursts (e.g., 4U 1722-30 and 4U 1812-12)
are of order 5 to 10 days \citep[e.g.,][]{zan07}, implying required
exposure times of $\sim 500\textrm{ ks}$ if one were to plan a
dedicated observation.  There is often some predictability to when an
X-ray burst will occur \citep[e.g.,][]{ube99,gal08} and with careful
planning less demanding exposure times (say by a factor of two) may be
feasible.

\acknowledgements 

We are grateful to David Ballantyne for providing us with his
reflection models, Craig Markwardt for guidance in the PCA background
calculation, Duncan Galloway for pointing out a second RXTE burst from
4U 1722-30, Erik Kuulkers for discovering the rich RXTE burst from 4U
0614+09, Nikolai Shaposhnikov for calling our attention to background
issues with the 4U 1722-30 data, and thank Lars Bildsten, Phil Chang,
Duncan Galloway, Alexander Heger, Jelle Kaastra, Laurens Keek, Frits
Paerels, Tod Strohmayer and Eliot Quataert for helpful discussions.

\bibliographystyle{aa} 
\bibliography{13952}

\appendix
\normalsize 
\section{PCA spectral data preparation and analysis method}
\label{sec:app3bursts}

Details about the PCA data collecting modes, analysis settings, and
time and spectral resolution are provided in
Table~\ref{tabobspca}. For 4U 0614+09 and 4U 1722-30 data were
obtained in the burst catcher mode. For 4U 1820-30 standard-2 data
were obtained.  The PCA spectra were accumulated from all active
PCUs. While the spectra for 4U 1820-30 were drawn from only the upper
xenon layer (`LR1'), those of the other two bursts were from all
layers (the `burst catcher' mode does not allow separation of PCUs or
layers).

A varying time resolution was employed in the time-resolved
spectroscopy, based on the resolution available, the statistical
quality of the data and the changes in the spectrum.  The spectra were
rebinned so as to have at least 15 photons per bin in order to ensure
applicability of the $\chi^2$ statistic. Contributions from particle
and cosmic backgrounds, as predicted through version 3.6 of the tool
{\it pcabackest}, were subtracted.

\renewcommand{\tabcolsep}{5mm}
\begin{table*}
\caption[]{Details of the data (analysis) for the 3 PCA superexpansion
  bursts. See also Figs.~\ref{figedge}, \ref{figsp}, and
  \ref{figdev}.\label{tabobspca}}
\begin{center}
\begin{tabular}{llll}
\hline\hline
  & 4U 0614+09 & 4U 1722-30 & 4U 1820-30$^1$ \\
\hline
Start time in spacecraft clock seconds$^2$  & 223941163.45 & 90054034.88 & 179459214.38 \\
UTC             & 4-Feb-01 & 8-Nov-96 & 9-Sep-99 \\
                & 21:52:34 & 7:00:31 & 1:47:58 \\
Active PCUs & 0,1,2,3 & 0,1,2,3,4 & 0,2,3 \\
Time interval of superexpansion$^3$ & 0.04-1.2 & 3.63-5.26 & 15-17 \\
End time of moderate expansion$^3$  & 89 & 23 & 1400 \\
Time interval with available burst catcher mode data$^3$ & 0.93/35.93 & -0.25/3.25 & n/a \\
 & & 7.75/97.50 & \\
Time/channel resolution burst catcher mode data & 8ms/64 & 2ms/64 & n/a \\
Time/channel resolution science event mode data & 125~$\mu$s/64 & 16~$\mu$s/64 & n/a \\
Time/channel resolution resolution standard-2 data & 16~s/129 & 16~s/129 & 16~s/129 \\
Time interval of initial spectral study (Fig.~\ref{figdev})$^3$ & 7.0-23.0 & -- & 325.1-1861.1 \\
Time interval of `pre-burst' spectrum$^3$& -740/-100 & +5760/+9220$^4$ & -3140/-2320\\
\hline\hline
\end{tabular}
\end{center}

$^1$4U 1820-30 data is accumulated only for the upper xenon layers;
$^2$ In spacecraft clock seconds (i.e., seconds since Jan. 1, 1994,
0:00:00.00 UTC); $^3$ Time in seconds since start time; $^4$For
  4U 1722-30 data were taken {\em after} the burst because pre-burst
  data are contaminated by a recent emergence of the spacecraft from
  the South Atlantic Geomagnetic Anomaly.
\end{table*}

Response matrices were generated with version 11.7 of the tool {\it
  pcarmf}.  A systematic error of 1.5\% was assumed per photon energy
channel. The instrument team recently
suggested\footnote{http://www.universe.nasa.gov/xrays/programs/rxte/pca/
  doc/rmf/pcarmf-11.7/} that the systematic error should be set to
0.5\%. While our fitted parameter values do not change if we adopt
their prescription, doing so increases the $\chi^2$ values somewhat.
Perhaps the larger systematic uncertainty is due to a spectrum (a
  few-keV black body) that is different from that of the Crab source
  (a power law with a photon index of 2.1) which was solely used for
  the calibration; this should be investigated further.

Burst spectra were corrected for detector dead time following the
procedure described in the RXTE Cook
Book\footnote{http://heasarc.gsfc.nasa.gov/docs/xte/recipes/pca\_deadtime.html}.
The time resolution for such a correction is limited to 0.125 s, so
interpolations were necessary for the precursor in 4U 0614+09.

The burst spectra were compared with models between 3 keV and 30 keV
using {\em XSPEC} version 12.5.0 \citep{arn96}. The 3 keV threshold
signifies the photon energy below which the instrument response is not
well calibrated. The upper threshold signifies the photon energy above
which the burst emission is expected to be negligible. At those
energies, the spectrum is expected to be dominated by the accretion
flux. Thus, some handle is obtained on the accretion emission.

Accounting for the changing accretion emission is not trivial. In
super-Eddington bursts, one cannot simply subtract the spectrum
accumulated before the burst because the super-Eddington flux and
expanding photosphere can affect the accretion disk
\citep[e.g.,][]{bal04}. Indeed, as shown in the main text, the
observed flux drops below the pre-burst (i.e., accretion) flux during
the superexpansion phase. The flux during the moderate expansion phase
may also affect the accretion. It is difficult to differentiate
accretion flux from burst flux, because the burst emission dominates
over the most sensitive part of the bandpass. Only at the highest
energies does the accretion emission dominate, but the statistical
quality is insufficient to infer a spectral shape. We assume that the
spectral shape of the accretion emission remains, throughout the
burst, identical to that of the pre-burst emission and fit its
(`accretion') normalization. When possible we determined an
empirical 3-30 keV spectral shape from data in a $\sim$1000~s time
interval prior to the burst (this is not always exactly 1000~s since
there is not always data available for a continuous 1000~s interval,
see Table~\ref{tabobspca}). The data from 4U 0614+09 are consistent
with a power law, and the data from 4U 1722-30 and 4U 1820-30 are
consistent with a comptonized spectrum, all with low-energy
absorption.  This accretion model was incorporated in to our full
spectral model of the data.

Quoted errors on spectral parameters are single-parameter 1$\sigma$
values.

\section{PCA spectral model}
\label{app:model}

\begin{figure*}[t]
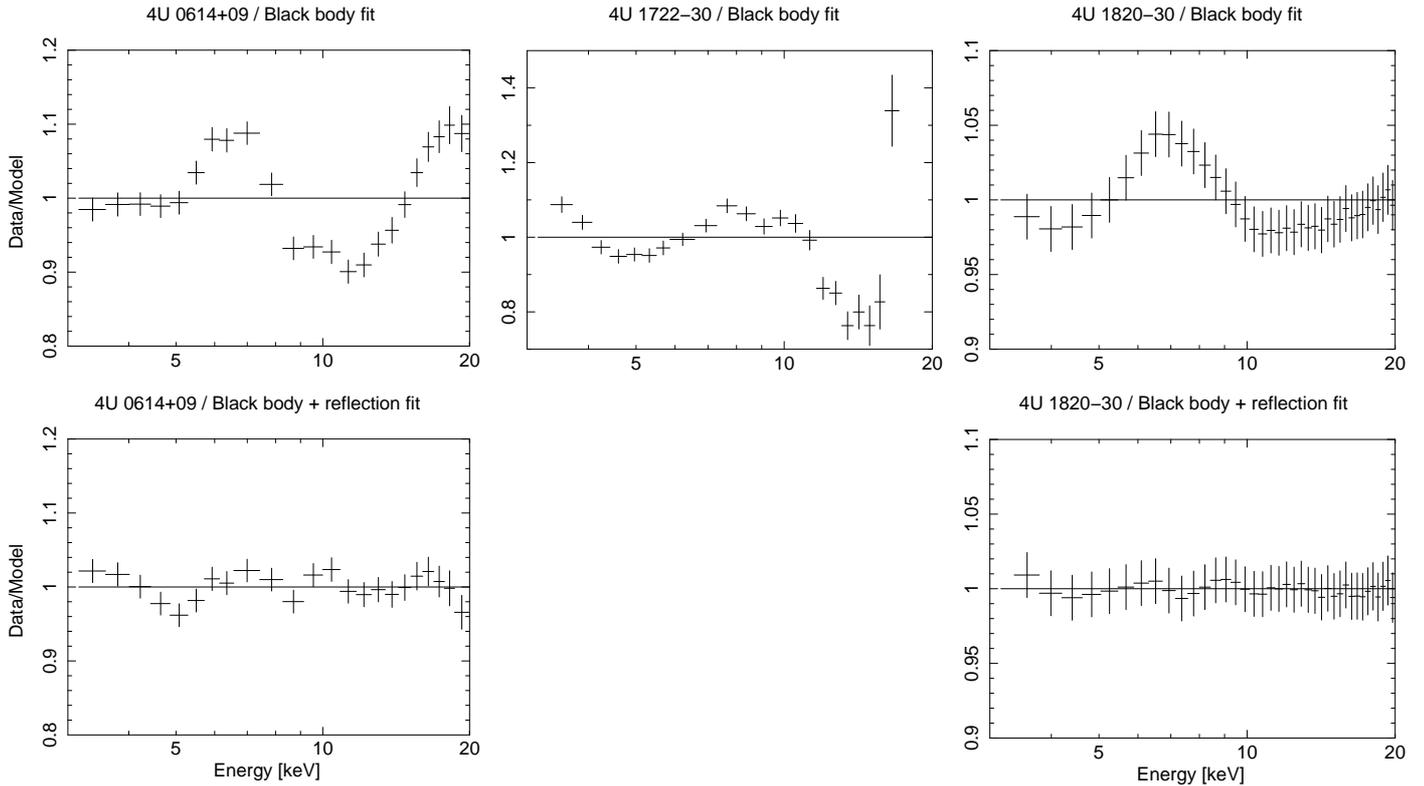


\renewcommand{\tabcolsep}{1mm}
\begin{tabular}{ccc}
\includegraphics[height=0.70\columnwidth,angle=270]{13952fa1a.ps} & \includegraphics[height=0.655\columnwidth,angle=270]{13952fa1b.ps} & \includegraphics[height=0.655\columnwidth,angle=270]{13952fa1c.ps} \\
\includegraphics[height=0.70\columnwidth,angle=270]{13952fa1d.ps} & & \includegraphics[height=0.655\columnwidth,angle=270]{13952fa1e.ps} \\
\end{tabular}

\caption{Spectral fits to the PCA bursts from 4U 0614+09 ({\em left
    plots}), 4U 1722-30 ({\em middle plot}) and 4U 1820-30 ({\em right
    plots}) during time intervals when the absorbed black body plus
  reflection model provides a good fit to the data (see
  Fig. \ref{figedge} for example time intervals when the black body
  plus reflection model fails to provide a good fit to these
  bursts). The upper row shows the fractional deviations for a model
  that includes the absorbed black body component but not the
  reflection or absorption edge components.  The lower row shows the
  deviations for a model that includes the absorbed black body and
  reflection component but not the absorption edge component. 4U
  1722-30 is not shown because at that time does a reflection
  component explain the deviations. The spectra are taken during the
  moderate expansion. The error bars include the 1.5\% systematic
  error.\label{figdev}}
\end{figure*}

To understand the X-ray burst emission process, we first accumulated
an optimum-quality spectrum for each burst by identifying long data
stretches when the spectral shape, as measured by the flux ratio
between two photon energy ranges, remains approximately constant.  The
times over which we accumulated data for this initial spectral study
are given in Table~\ref{tabobspca}. We then tested whether the
resulting spectra are consistent with the canonical burst model,
consisting of black body radiation (leaving free the temperature and
normalization), low-energy absorption by cold gas of cosmic abundances
(following the model for atomic cross sections and cosmic abundances
of \citealt{mor83}; leaving free the equivalent H-atom column density
$N_{\rm H}$) and accretion emission (leaving free the accretion
normalization; see above).  The fractional deviations of the spectra
with respect to this model are shown in the upper row of
Fig.~\ref{figdev}.  Not shown are the results between 20 and 30 keV
because of the large error bars in that bandpass. None of the spectra
fit the model. The shapes of the deviations are generally similar with
a maximum at 6 to 7 keV and a minimum between 10 and 13 keV. The
amplitude of the deviations of the different bursts are similar and
far larger than instrumental uncertainties (0.5\% to 1.5\% per
channel).

One approach to modeling the deviations is to include both a broad
Gaussian emission line centered between 6 and 7 keV and an absorption
edge close to 10 keV. A more physically motivated approach is to
account for scattering (`reflection') and relativistic blurring of the
burst's black body radiation by a hot accretion disk
\citep{bal04,bal04b}. \cite{bal04} used such a reflection model to fit
the deviations in the 4U 1820-30 superburst. Their reflection model
accounts for reprocessing of thermal photons from the neutron star by
the accretion disk (including comptonization, fluorescence,
recombination, and photo-electric absorption), which is allowed to
come into thermal and ionization balance. The model includes a number
of ionization stages of C, N, O, Mg, Si and Fe, and the parameters
include the bolometric flux of the reflection, the temperature of the
black body, metal abundance and the ionization degree $\xi=4\pi F_{\rm
  X}/n_{\rm H}$ of the scattering medium, with $F_{\rm X}$ the
bolometric flux of the direct black body radiation and $n_{\rm H}$ the
hydrogen number density. See  \cite{bal04b} for further details
of this model.

We used the \cite{bal04b} reflection model. In our implementation, we
assume solar abundances and fit for two of the reflection parameters:
$\xi$ and the bolometric flux. Assuming solar abundances is only
relevant to Fe since it is the only element in the model with
spectral features in the PCA bandpass. The temperature is given by
that of the black body component. Note that this model is quite
different from reflection by a power law as commonly applied in AGN
and Galactic black hole spectra. In bursts, the radiant energy per
frequency decade (the `$\nu F_\nu$' spectrum) peaks for black bodies
at photon energies where photoelectric absorption by metals is very
important, unlike for power law emission in AGN and Galactic black
holes \citep{bal04b}.

The reflection model also includes a convolution function for
relativistic blurring by the accretion disk following \cite{fab89}. It
is parametrized by the power law index for the radial dependence of
the emissivity (set to -3), the inner disk radius (set to 10G$M/c^2$
with $M$ the neutron star mass), the outer disk radius (set to
200G$M/c^2$) and the inclination angle (set to 30$^{\rm o}$). Whenever
these parameters were left free during fitting, the improvement in the
fit was marginal. For practical purposes they were, therefore, kept
fixed.

The complete applied model can  be written in XSPEC as: {\it
  constant*(wabs$_1$*(bbodyrad + rdblur*atable\{reflection\}) +
  wabs$_2$*comptt/po)} with {\it constant} the detector dead time
factor, {\it wabs$_1$} and {\it wabs$_2$} low-energy absorption
(different for the two instances in this formula), {\it bbodyrad} the
Planck function, {\it rdblur} the relativistic blurring function, {\it
  reflection} the model for the reflection of the Planck function
against the accretion disk following \cite{bal04b}, {\it comptt} a
comptonization component following \cite{tit94} and {\it po} a power
law function.  The latter two components are representative of the
non-burst emission. The model has six free parameters: $N_{\rm H}$,
k$T_{\rm bb}$, R$_{\rm bb}$, the flux and $\xi$ of the reflection
component and the accretion normalization of {\it comptt} or {\it
  po} expressed relative to the pre-burst value.

Including reflection in the model is highly successful in reducing the
deviations in the spectra of 4U 0614+09 and 4U 1820-30 for the
particular time interval specified in Table~\ref{tabobspca} (the
``time interval of initial spectral study"). We find $\chi^2_\nu$=2.3
and 0.75, respectively \citep[the value for 4U 1820-30 is consistent
  with][]{bal04}. The remaining deviations are shown in the lower row
of Fig.~\ref{figdev}. The black body plus reflection model does not,
however, fit all the spectra of 4U 0614+09 and 4U 1820-30 (see
\S~\ref{sec:abs_edges_3PCA}). Furthermore, it almost never provides a
good fit to the spectra of 4U 1722-30 (the middle-right panel in
Fig. \ref{figedge} is representative of the typical spectral
deviations found at all times in black body plus reflection model fits
to the 4U 1722-30 spectra).

Even when the black body plus reflection model provides a good fit,
the flux of the reflection component, a free parameter, is often
higher than that of the direct black body component. This is due to
the fact that the fitting procedure primarily tries to fit the narrow
features -- the broad emission line and the absorption edge. Since the
continua of the reflection and direct black body components are very
similar in the PCA bandpass, their contributions are difficult to
disentangle. High reflection fluxes were also found by \cite{bal04} in
the late-time burst spectra of 4U 1820-30 (see their Fig.~1). As a
result of a high reflection flux, the flux of the direct black body
component becomes so low that the black body radius becomes
substantially lower than 10 km. This is physically
unrealistic. Without a proper model for the geometry of the accretion
disk and independent measurements of the iron abundance in the donor
star, it is impossible to meaningfully constrain the reflection flux
in an independent manner.

We calculated bolometric luminosities and emission region radii by
fitting the spectra without the reflection (or edge) component. After
fitting, the absorption and the accretion normalization was set to
zero, the deadtime correction set to 1, and the response matrix in
{\em XSPEC} set to a dummy value. The flux could then be estimated in
a broader energy range than the PCA bandpass. The unabsorbed flux was
calculated between 0.05 and 30 keV, essentially a bolometric bandpass
for a black body with a temperature between 0.5 and 3 keV. Since the
fits are often formally unacceptable, no errors were calculated.

We tried several other functions to model the spectra. These include
combinations of two different black bodies, combinations of one black
body with a power law (with a free spectral index), gauss function or
relativistic iron line \citep[following][model {\it laor} in {\it
    XSPEC}]{laor}, and single reflection models of a power law against
a neutral or ionized disk \citep[following][models {\it pexrav/pexriv}
  in {\it XSPEC}]{zdiar}. While some of these models fit some of the
spectra reasonably well, they only do so over limited time intervals
in any given burst. A broken power law fits the data with the same
goodness ($\chi^2_\nu$) at any time, but there is no physical basis
for such a model.

\section{Ruling out alternative explanations for the lack of correlation
between $t_{\rm se}$ and $\tau_{\rm decay}$}
\label{app:unlikely_tse_explanations}

Here we first consider the possibility that the transition
between the superexpansion phase and the moderate expansion phase is
due not to a shell ejection, as suggested in
\S~\ref{sec:tse_discussion}, but rather to a transition in the
luminosity at the NS surface. In particular, if the luminosity exceeds
not only the \emph{local} Eddington luminosity, $L_{\rm Edd}\equiv
4\pi G M c / \kappa_0$, but also the Eddington luminosity at infinity,
$L_{\rm Edd, \infty}\simeq L_{\rm Edd} (1+z)^2$, a strong
radiation-driven wind will form ($\kappa_0=0.2 \textrm{ cm$^2$
  g}^{-1}$ is the hydrogen-free electron scattering opacity). Such a
wind can potentially drive the photosphere to very large radii ($\ga
100 \textrm{ km}$). If, however, $L_{\rm Edd} < L < L_{\rm Edd,
  \infty}$, a comparatively mild quasi-static expansion is sufficient
to reduce the luminosity at the photosphere below $L_{\rm Edd}$
\citep{han82, Paczynski:83, pac86}. In this case, the PRE will be less
extreme. This suggests that perhaps the large radius expansion factors
reached during the superexpansion occur because initially $L>L_{\rm
  Edd, \infty}$. Then, after a time $t_{\rm se} \approx \textrm{
  seconds}$, the luminosity decreases to $L_{\rm Edd} < L < L_{\rm
  Edd, \infty}$ and the photosphere retreats closer to the NS surface,
marking the start of the moderate expansion phase. This phase
continues for a duration of order the cooling time after which the
luminosity decreases to $L<L_{\rm Edd}$ and the photosphere settles on
to the NS surface.

To test this idea, we carry out time-dependent cooling calculations
for a range of ignition depths. To solve the heat transport equation
we use the technique described in \citet{wei07}; \citet{cum04} carry
out similar calculations for superbursts. We assume that the
atmosphere is radiative and initially at its peak temperature, in
accord with models of the rise of X-ray bursts (see e.g.,
\citealt{Cumming:00, wei06a}). We assume a peak temperature at the
base of the ignition layer $y_{\rm base}$ of $T_{\rm base, initial} =
f T_{\rm max}$, where $f$ is factor less than but of order unity and
$T_{\rm max}=(3 g y_{\rm base} / a)^{1/4}$ is the critical temperature
at which radiation pressure completely dominates ($g=(1+z)GM/R^2$ is
the gravitational acceleration and $a$ is the radiation constant). We
chose the value of $f$ by solving for $T_{\rm base, initial}$ such
that the fluence (the cooling curve integrated over the duration of
the burst) equals the total nuclear energy release $E=4\pi R^2 y_{\rm
  base} Q_{\rm nuc} / (1+z)$ assuming $Q_{\rm nuc}=1.6 \textrm{ MeV
  nucleon}^{-1}$, corresponding to helium burning completely to
nickel. Although our conclusions are not sensitive to the exact value
of $f$, for reference, we find $f\approx 0.8-0.9$.

\begin{figure}[t]
\includegraphics[bb= 20 0 0 590, height=9.0cm,angle=0]{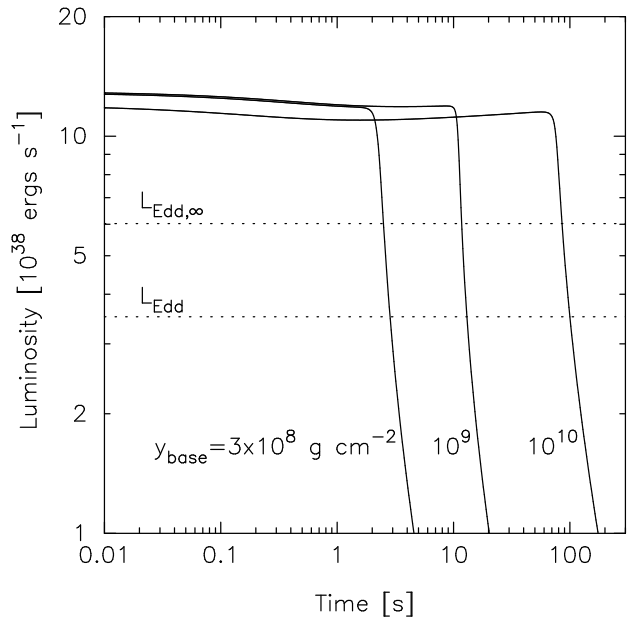}
\caption{Model calculations of the radiative cooling following the
  burst rise. The solid lines show the luminosity $L$ for ignition
  depths $y_{\rm base}=3\times10^8, 10^9,$ and $10^{10}\textrm{ g
    cm}^{-2}$ with $f=0.88, 0.85,$ and $0.78$, respectively. The
  values for $f$ were chosen such that the fluence equals the nuclear
  energy released in burning the entire He column to Ni. The Eddington
  limit at the NS surface and at infinity (assuming $z=0.31$ and
  $M=1.4$~M$_\odot$) are drawn as dotted lines.\label{figmodels}}
\end{figure}

In Fig.~\ref{figmodels} we show time profiles of the calculated
luminosity $L$ for three ignition depths $y_{\rm base}$. The dotted
lines mark $L_{\rm Edd}$ and $L_{\rm Edd, \infty}$ for a 1.4~M$_\odot$
NS and an opacity $\kappa_0=0.2 \textrm{ cm$^2$ g$^{-1}$}$. An
observer would see an Eddington-limited light curve where $L>L_{\rm
  Edd}$ (the excess luminosity is used to drive the outward expansion
of the overlying layers) and a light curve that follows the plotted
luminosity where $L<L_{\rm Edd}$.

As the figure shows, the profiles all have the same shape regardless
of $y_{\rm base}$ and $f$: an early flat phase followed by a break and
a steep decay. The factor $f$ primarily determines the overall
vertical scale while $y_{\rm base}$ determines the thermal time of the
ignition layer and therefore the location of the break. The key result
is that the early-time light curve is very flat, barely decreasing
until after the break, at which point it rapidly falls below $L_{\rm
  Edd}$. Such a light curve cannot explain the features of the
superexpansion as there is no second-long, depth-independent phase
during which $L>L_{\rm Edd, \infty}$.
 
The light curve is flat because we assumed that the initial profile
was the radiative profile ($T\propto y^n$ with $n\approx 1/4$) and
therefore $L\propto T^4/y$ is constant until the thermal wave reaches
$y_{\rm base}$ and the ignition layer begins to cool. If we had
assumed a shallower initial profile, the early-time flux would
decrease with time (see, e.g., the superburst calculations of
\citealt{cum04} who assume an initial profile with $n=1/8$
corresponding to an isobaric deflagration).  However, absent any
strong motivation for assuming a non-radiative initial profile, we
conclude that a variation in the early time $L$ is unlikely to
explain the lack of correlation between $t_{\rm se}$ and $\tau_{\rm
  decay}$.
  
Since the superexpansion phase always occurs within seconds of burst
onset, perhaps it is associated not with post-peak cooling, as 
assumed above, but rather with the physics of the burning during the
burst rise. Due to their larger ignition depths and helium mass
fractions, the peak energy generation rate in superexpansion bursts
can be considerably higher than in ordinary short-duration bursts. If
the ignition is deep enough, the heating time due to burning $t_{\rm
  heat} = (d\ln T_b/dt)^{-1}$ can become shorter than the local
dynamical time $t_{\rm dyn} \approx 10^{-6}\textrm{ s}$. Such hydrodynamic
burning could drive strong shocks through the upper layers of the
atmosphere \citep{Zingale:01, wei06b}. However, when we solve the rise
using the prescription described in \citet{wei06a} assuming an
initially pure He layer, we find that $t_{\rm heat} > t_{\rm dyn}$
unless $y_{\rm base} \ga 10^{11} \textrm{ g cm}^{-2}$. This suggests
that strong shocks do not form during the rise of superexpansion
bursts as they ignite at $y_{\rm base} \la 10^{10} \textrm{ g
  cm}^{-2}$.

\end{document}